\documentclass{article}
\usepackage{amssymb}
\usepackage{amsmath}
\usepackage[dvips]{graphicx}
\usepackage{fleqn,ams}
\usepackage{verbatim}
\usepackage[cp1251]{inputenc}
\usepackage[dvips]{graphicx}
\begin{document}
\begin{center}
{\bf\Large Dynamic model of spherical perturbations in the
Friedman universe. III. Automodel solutions.} \\[12pt]
Yu.G. Ignatyev, N. Elmakhi\\
Tatar State Humanitarity Pedagocical University\\ 1 Mezhlauk St.,
Kazan 420021, Russia
\end{center}

\begin{abstract}
A class of exact spherically symmetric perturbations of retarding
automodel solutions linearized around Friedman background of
Einstein equations for an ideal fluid with an arbitrary barotrope
value is obtained and investigated.
\end{abstract}
\section{Introduction}
In the previous works of the authors [1,2] a class of exact
retarding solutions for linear spherical perturbations of the
Friedman universe  with an ultrarelativistic equation of state of
the ideal fluid filling it, corresponding to the central singular
source presence and having the sight of polinoms by radial
variable, was obtained. At that it was noted, by zero boarding
conditions at the sound horizon for the   $C^{1} $class metrics
perturbations the energy dense perturbations have the first genus
break at the sound horizon.  In this paper we shall study the
retarding solutions in detail by extending the investigations
range to the equations of state of the fluid with an arbitrary
barotrope coefficient $\kappa $.

So we shall investigate the retarding solutions of the
evolutionary equation for the spherical perturbations [1] (84):

\begin{equation} \label{GrindEQ__1_}
\ddot{\Psi }+\frac{2}{\eta } \dot{\Psi }-\frac{6(1+\kappa )}{(1+3\kappa )^{2} } \frac{\Psi }{\eta ^{2} } -\kappa \Psi ''=0.
\end{equation}
with boarding conditions at the sound horizon corresponding to the zero values of the metrics perturbations and its first radial derivatives

\begin{equation} \label{GrindEQ__2_}
\Sigma _{0} :\quad r=\sqrt{\kappa } \eta .
\end{equation}

\begin{equation} \label{GrindEQ__3_}
\mathop{\left. \Psi (r,\eta )\right|}\nolimits_{r\sqrt{\kappa } \eta } =\mu (\eta );\quad \quad \mathop{\left. \Psi '(r,\eta )\right|}\nolimits_{r\sqrt{\kappa } \eta } =0,
\end{equation}
where $\mu (\eta )$ is the central source singular mass, so that  the metrics component perturbation $g_{44} $ is equal to

\begin{equation} \label{GrindEQ__4_}
\delta g_{44} =a^{2} (\eta )\delta \nu ,\quad \delta \nu =2\frac{\Phi (r,\eta )}{ar} =2\frac{\Psi (r,\eta )-\mu (\eta )}{ar} ,
\end{equation}
moreover the function $\Psi (r,\eta )$ corresponds to the nonsingular part of the potential:

\begin{equation} \label{GrindEQ__5_}
\Psi (0,\eta )=0.
\end{equation}
Temporal mass evolution is described by the equation

\begin{equation} \label{GrindEQ__6_}
\ddot{\mu }+\frac{2}{\eta } \dot{\mu }-\frac{6(1+\kappa )}{(1+3\kappa )^{2} } \frac{\mu }{\eta ^{2} } =0,
\end{equation}
which has the solution

\begin{equation} \label{GrindEQ__7_}
\mu =\mu _{+} \eta ^{{\tfrac{2}{1+3\kappa }} } +\mu _{-} \eta ^{-{\tfrac{3(1+\kappa )}{1+3\kappa }} } ,\quad \quad (1+\kappa )\ne 0;
\end{equation}
(for detail see the previous works [1,2]).

\noindent The potential function $\Psi (r,\eta )$ and the scalar $\mu (\eta )$ completely determine the energy density perturbations and the fluid velocity

\begin{equation} \label{GrindEQ__8_}
\frac{\delta \varepsilon }{\varepsilon _{0} } =-\frac{1}{4\pi ra^{3} \varepsilon _{0} } \left(3\frac{\dot{a}}{a} \dot{\Phi }-\Psi ''\right),
\end{equation}

\begin{equation} \label{GrindEQ__9_}
(1+\kappa )v=-\frac{1}{4\pi ra^{3} \varepsilon _{0} } \frac{\partial }{\partial r} \frac{\dot{\Phi }}{r} ,
\end{equation}
where $\varepsilon _{0} (\eta )$ is a non-perturbed energy density of the Friedman universe

\begin{equation} \label{GrindEQ__10_}
\varepsilon _{0} \sim \eta ^{-{\tfrac{6(1+\kappa )}{1+3\kappa }} } ;\quad a\sim \eta ^{{\tfrac{2}{1+3\kappa }} } ;\quad \varepsilon _{0} a^{3} \sim \eta ^{-{\tfrac{6\kappa }{1+3\kappa }} } .
\end{equation}

\section{Automodel solutions}

\subsection{A general automodel solution}
So we shall look for solutions of the evolutionary equations for
the perturbations with zero boarding conditions at the zero sound
horizon \eqref{GrindEQ__2_}. Supposing that in
\eqref{GrindEQ__1_}:

\begin{equation} \label{GrindEQ__11_}
\Psi (r,\eta )=\eta ^{\alpha } G_{\alpha } (z),
\end{equation}
where
\begin{equation} \label{GrindEQ__12_}
z=\frac{r}{\sqrt{\kappa } \eta } ,
\end{equation}
we come to the \textit{automodel solutions }class and get a common
differential equation for the function $G_{\alpha } (z)$:

\begin{equation} \label{GrindEQ__13_}
(1-z^{2} )\mathop{G''}\nolimits_{\alpha } (z)+2\alpha z\mathop{G'}\nolimits_{\alpha } (z)+
\left[\frac{6(1+\kappa )}{(1+3\kappa )^{2} } -\alpha (1+\alpha )\right]G(z)_{\alpha } =0.
\end{equation}
The potential function $\Psi (r,\eta )$ must be a combination of the private solutions

\begin{equation} \label{GrindEQ__14_}
\Psi (r,\eta )=\sum _{\alpha } \, \eta ^{\alpha } G_{\alpha } (z).
\end{equation}
From \eqref{GrindEQ__3_} and \eqref{GrindEQ__7_} it follows that
by $\alpha \ne 0$ this combination can have  two members only

\begin{equation} \label{GrindEQ__15_}
\Psi (r,\eta )=G_{+} (z)\eta ^{{\tfrac{2}{1+3\kappa }} } +
G_{-} (z)\eta ^{-{\tfrac{3(1+\kappa )}{1+3\kappa }} } ,
\end{equation}
at that in consequence of the boarding conditions
\eqref{GrindEQ__3_} the functions $G_{\pm } (z)$ must satisfy the
following boarding conditions

\begin{equation} \label{GrindEQ__16_}
G_{\pm } (1)=\mu _{\pm } ;\quad \quad \mathop{G'}\nolimits_{\pm } (1)=0.
\end{equation}
In particular from \eqref{GrindEQ__3_} and \eqref{GrindEQ__8_} it follows immediately that in the case $\alpha =0$ the zero mass of the singular source corresponds to the pointed out class of solutions.

\noindent At the parameter $\alpha $ arbitrary values  the general solution of the linear differential equation \eqref{GrindEQ__13_} is expressed by the Legendre functions, $P_{\nu }^{\mu } (z)$, and the adjoint Legendre functions  $Q_{\nu }^{\mu } $\footnote{  For example, see [7] and [8].}:

\begin{equation} \label{GrindEQ__17_}
G(z)=\mathop{\left(\frac{1-z}{1+z} \right)}\nolimits^{{\tfrac{\alpha +1}{2}} } \left[C_{1} P_{{\tfrac{3(1-\kappa )}{2(1+\kappa )}} }^{\alpha +1} (z)+C_{2} Q_{{\tfrac{3(1-\kappa )}{2(1+\kappa )}} }^{\alpha +1} (z)\right].
\end{equation}

\subsection{ An auto model solution with the particle-like
source ($\mu \ne 0$)} The mentioned above is just for the formal
general solution \eqref{GrindEQ__17_} at an arbitrary value of the
parameter $\alpha $. However,  in our particular case
\eqref{GrindEQ__15_} the values of the parameter $\alpha $:

\begin{equation} \label{GrindEQ__18_}
\alpha =\left(\frac{2}{1+3\kappa } ,-\frac{3(1+\kappa )}{1+3\kappa } \right)
\end{equation}
are the quadratic equation roots simultaneously

\begin{equation} \label{GrindEQ__19_}
\frac{6(1+\kappa )}{(1+3\kappa )^{2} } -\alpha (1+\alpha )=0.
\end{equation}
Therefore in this case the equation \eqref{GrindEQ__13_} degenerates into a simpler one

\begin{equation} \label{GrindEQ__20_}
(1-z^{2} )\mathop{G''}\nolimits_{\alpha } (z)+2\alpha z\mathop{G'}\nolimits_{\alpha } (z)=0,
\end{equation}
However, integrating it we obtain the following

\begin{equation} \label{GrindEQ__21_}
\mathop{G'}\nolimits_{\alpha } (z)=C_{1} (1-z^{2} )^{\alpha } .
\end{equation}
Comparing the second boarding condition \eqref{GrindEQ__16_} with the expression \eqref{GrindEQ__21_} we see that for the fulfillment of the zero conditions for the first derivative of the potential at the zero sound horizon it is necessary  that $\alpha >0$, that is  The particle-like source mass is growing by the time and is equal to zero at the time moment $\eta =0$; thus, =1 to provide a smooth sewing together of the solution with the Friedman one at the zero sound horizon it is necessary that

\begin{equation} \label{GrindEQ__22_}
\mu _{-} =0.
\end{equation}
In the case of  $\alpha >0$in consequence of the relation \eqref{GrindEQ__21_} the second boarding condition \eqref{GrindEQ__16_} is fulfilled by fulfilling the first one automatically, thus, the condition $\alpha >0$ fulfillment provides a smooth sewing together of the solution in the $C^{1} $ class at the sound horizon.

\noindent Now integrating the equation \eqref{GrindEQ__21_} formally and considering the conditions at the beginning of the coordinates  \eqref{GrindEQ__5_}, according to which

\begin{equation} \label{GrindEQ__23_}
G(0)=0,
\end{equation}
we find its formal solution within the whole interval of the values $r=[0,+\infty )$:

\begin{equation} \label{GrindEQ__24_}
\hspace{-20pt}G(\kappa ,z)= C_{1} \left\{
\begin{array}{l}
zF\left(\frac{1}{2} ,-\frac{2}{1+3\kappa
} ,\frac{3}{2} ,z^{2} \right),\quad (z\le 1); \\[12pt]
{\displaystyle \frac{\sqrt{\pi } \Gamma \left({\tfrac{2}{1+3\kappa
}} +1\right)}{2\Gamma \left({\tfrac{2}{1+3\kappa }}
+{\tfrac{3}{2}} \right)} + \int _{0}^{\ln (z+\sqrt{z^{2} -1} )} \,
^{{\tfrac{4}{1+3\kappa }} +1} xdx,}\quad (z>1),\end{array}\right.
\end{equation}
where $F(a,b,c,x)$ is a hyper-geometrical function (for example ,see [7]):

\begin{eqnarray} \label{GrindEQ__25_}
F(\alpha ,\beta ,\gamma ,z)= \frac{\Gamma (\gamma )}{\Gamma (\beta
)\Gamma (\gamma -\beta )} \int _{0}^{1} \, t^{\beta -1}
(1-t)^{\gamma -\beta -1} (1-tz)^{-\alpha } ,\\[12pt]
\nonumber\Re (\gamma )>\Re (\beta )>0;\; |{\rm arg}(1-{\rm
z})|<\pi .
\end{eqnarray}
At that the useful limitary relation is just [7]:

\begin{equation} \label{GrindEQ__26_}
\mathop{\lim }\limits_{z\to 1-0} F(\alpha ,\beta ,\gamma
,z)=\frac{\Gamma (\gamma )\Gamma (\gamma -\alpha -\beta )}{\Gamma
(\gamma -\alpha )\Gamma (\gamma -\beta )} ,\quad \Re (\gamma
-\alpha -\beta )>0;
\end{equation}
where $\Gamma (x)$ is $\Gamma $-function. In Fig. 1 the solutions
\eqref{GrindEQ__24_} for values series of the barotrope
coefficient are shown.

\begin{tabular}{l}
\includegraphics[width=105.7mm, height=76.4mm]{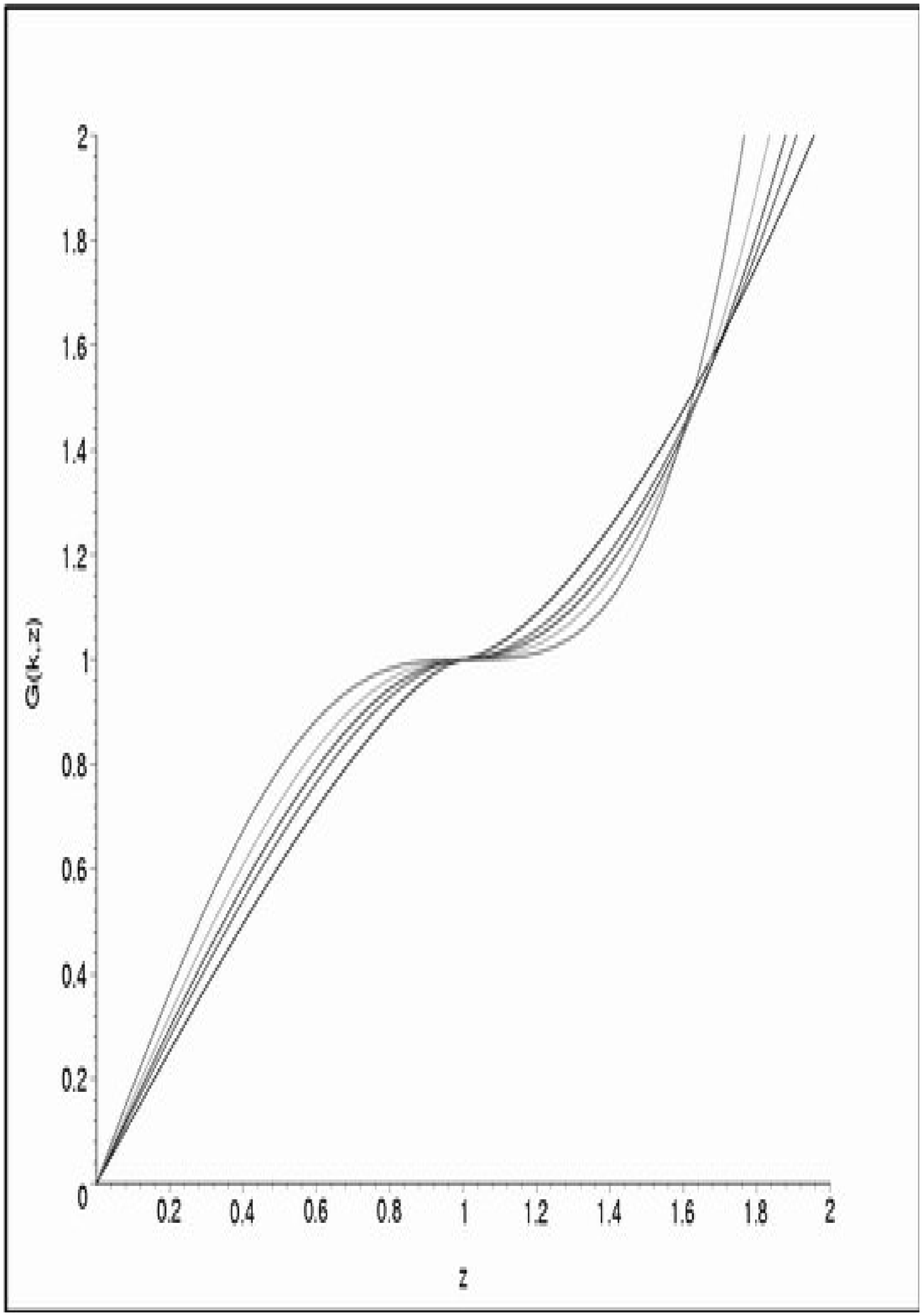} \\
\parbox{105.7mm}{\textbf{Fig. 1. Normalized function} $G(k,z)$ calculated with
the formula \eqref{GrindEQ__24_} at the value \eqref{GrindEQ__27_}
of the constant $C_{1} $ and $\mu =1$. In the left part of the
figure from bottom up $\kappa =0;{\tfrac{1}{6}} ;{\tfrac{1}{3}}
;{\tfrac{2}{3}} ;1$.} \\
\end{tabular}
\vskip 12pt \hrule \vskip 12pt

In terms of the obtained solution \eqref{GrindEQ__24_} it is easy
to see $G(-z)=G(z)$, that is the function $\Psi (r,\eta )$ is  an
odd function of a radial variable really, it was shown in Ref.
[2]. Considering the first boundary condition
\eqref{GrindEQ__16_}, we find the constant $C_{1} $:

\begin{equation} \label{GrindEQ__27_}
C_{1} =\mu _{+} \frac{2\Gamma \left({\tfrac{2}{1+3\kappa }} +{\tfrac{3}{2}} \right)}{\sqrt{\pi } \Gamma \left({\tfrac{2}{1+3\kappa }} +1\right)}
\end{equation}
and thus, finally we obtain the automodel  $C^{1} $class solution corresponding to the zero boundary conditions at the sound horizon

\begin{eqnarray} \label{GrindEQ__28_}
\Psi (r,\eta ,\kappa )=\mu _{+} \eta ^{{\tfrac{2}{1+3\kappa }}
}\times\nonumber\hskip 6cm\\[12pt]
\left\{%
\begin{array}{ll}
{\displaystyle \frac{2\Gamma \left({\tfrac{2}{1+3\kappa }}
+{\tfrac{3}{2}} \right)}{\sqrt{\pi } \Gamma
\left({\tfrac{2}{1+3\kappa }} +1\right)} \frac{r}{\sqrt{\kappa }
\eta } F\left(\frac{1}{2} ,-\frac{2}{1+3\kappa } ,\frac{3}{2}
,\frac{r^{2} }{\kappa \eta
^{2} } \right)}, & (r\le \sqrt{\kappa } \eta );\\
1, & (r>\sqrt{\kappa } \eta ),
\end{array}
\right.
\end{eqnarray}
Taking into consideration the relation \eqref{GrindEQ__28_} according to  \eqref{GrindEQ__4_} we obtain the scalar function of the metrics perturbation $\delta \nu $:

\begin{eqnarray} \label{GrindEQ__29_}
\delta \nu (r,\eta )=-\frac{2\mu _{+} }{r}\times \hskip 7cm\nonumber\\
\left[1-\frac{2\Gamma \left({\tfrac{2}{1+3\kappa }}
+{\tfrac{3}{2}} \right)}{\sqrt{\pi } \Gamma
\left({\tfrac{2}{1+3\kappa }} +1\right)} \frac{r}{\sqrt{\kappa }
\eta } F\left(\frac{1}{2} ,-\frac{2}{1+3\kappa } ,\frac{3}{2}
,\frac{r^{2} }{\kappa \eta ^{2} } \right)\right]\chi (\sqrt{\kappa
} \eta -r),
\end{eqnarray}
where $\chi (x)$ is the Heavyside function:

\begin{equation} \label{GrindEQ__30_}
\chi (x)=\left\{0,\_ \_ x\le 0;1,\_ \_ x>0.\right.
\end{equation}

In the private values series of the barotrope coefficient $\kappa
$ the obtained solution is expressed in elementary functions
$\kappa =1/3$ the  ultra-relativistic fluid

\[G(1/3,z)=z-\frac{1}{3} z^{3} ;\]

\begin{equation} \label{GrindEQ__31_}
\Psi =\frac{3}{2} \mu _{+} \eta \left(z-\frac{1}{3} z^{3} \right);
\quad \delta \nu =-\frac{2\mu _{+} }{r} \left[1-\frac{3}{2}
\frac{r}{\sqrt{3} \eta } +\frac{1}{2} \mathop{\left(\frac{r}{\sqrt{3} \eta } \right)}\nolimits^{3} \right];
\end{equation}
Being auto-model the solution \eqref{GrindEQ__31_} coincides with
the general solution in the form of the degree series, earlier
obtained in the works [3,4,5,6], it confirms the correctness of
the proved in  [2] theorem about the retarding solution uniqueness
in the case of the ultra-relativistic state equation once more.
$\kappa =1$: fluid with extremely stiff state equation
\begin{eqnarray} \label{GrindEQ__32_}
G(1,z)=\frac{1}{2} z\sqrt{1-z^{2} } +\frac{1}{2} \arcsin
z;\nonumber\\ \Psi =\frac{2\mu _{+} }{\pi } \sqrt{\eta }
\left(\frac{r}{\eta } \sqrt{1-\frac{r^{2} }{\eta ^{2} } } +
2\arcsin \frac{r}{\eta } \right);\nonumber\\
\delta \nu =-\frac{2\mu _{+} }{r} \left[1-\frac{2}{\pi }
\left(\frac{r}{\eta } \sqrt{1-\frac{r^{2} }{\eta ^{2} } } +\arcsin
\frac{r}{\eta } \right)\right].
\end{eqnarray}

Note, the first derivative conversion by a radial variable at the
zero sound horizon in the obtained solutions at $\alpha >0$
guaranties the relation \eqref{GrindEQ__21_}.

\subsection{A solution without a particle-like source}
Now supposing that in \eqref{GrindEQ__11_} $\alpha =0$, we get the
following equation instead of  \eqref{GrindEQ__13_}

\begin{equation} \label{GrindEQ__33_}
(1-z^{2} )G''(z)+\frac{6(1+\kappa )}{(1+3\kappa )^{2} } G(z)=0.
\end{equation}
The general solution of this equation is a linear combination of the hyper-geometrical functions

\begin{eqnarray} \label{GrindEQ__34_}
G(\kappa ,z)=(1-z^{2} )\times\hskip 7cm\nonumber\\[12pt]
\left[zC_{1}F\left(\frac{3\kappa }{1+3\kappa } , \frac{5+9\kappa
}{2+6\kappa } ,\frac{3}{2} ,z^{2} \right)+C_{2}
F\left(\frac{2+3\kappa }{1+3\kappa } , \frac{3\kappa -1}{2+6\kappa
} ,\frac{1}{2} ,z^{2} \right)\right].
\end{eqnarray}
If the geometrical functions

\[F\left(\frac{3\kappa }{1+3\kappa } ,\frac{5+9\kappa }{2+6\kappa } ,\frac{3}{2} ,z^{2} \right)\quad and\quad F\left(\frac{2+3\kappa }{1+3\kappa } ,\frac{3\kappa -1}{2+6\kappa } ,\frac{1}{2} ,z^{2} \right)\]
remained final at the sound horizon $z=1$, the general solution \eqref{GrindEQ__33_} would automatically turn into zero at the sound horizon, then the zero boundary conditions at the sound horizon \eqref{GrindEQ__16_} would be in principle satisfied by the choice of constants in the general solution \eqref{GrindEQ__33_}. However the pointed out hyper-geometrical functions have some peculiarities at the sound horizon.

\noindent Really, as it is easy to see, the parameters $\alpha ,\beta ,\gamma $ of the hyper-geometrical functions of this linear combination satisfy the condition (See the hyper-geometrical function definition (26).)

\begin{equation} \label{GrindEQ__35_}
\alpha +\beta -\gamma =1,
\end{equation}
At that for the first member of this combination the additional condition is fulfilled

\begin{equation} \label{GrindEQ__36_}
\gamma _{1} -\beta _{1} =-\frac{1}{1+3\kappa } ,
\end{equation}
at the same time  for the second one it is

\begin{equation} \label{GrindEQ__37_}
\gamma _{2} -\beta _{2} =+\frac{1}{1+3\kappa } .
\end{equation}
In this case it  is necessary to use the functional relation for the hyper-geometrical function  [7], which conformably to \eqref{GrindEQ__35_} is given with the formula

\begin{equation} \label{GrindEQ__38_}
F(\alpha ,\beta ,\alpha +\beta -1,z)=\frac{1}{(1-z)} F(\alpha -1,\beta -1,\alpha +\beta -1)N
\end{equation}
Using this relation in the formula \eqref{GrindEQ__34_} let us rewrite the general solution in the form

\begin{eqnarray} \label{GrindEQ__39_}
G(\kappa ,z)=(1+z)\left[zC_{1} F\left(-\frac{1}{1+3\kappa }
,\frac{3(1+\kappa )}{2(1+3\kappa )} , \frac{1}{2} ,z^{2}
\right)\right.+\nonumber\\
\left.C_{2} F\left(\frac{1}{1+3\kappa } ,\frac{3(1+\kappa
)}{2(1+3\kappa )} ,-\frac{1}{2} ,z^{2} \right)\right].
\end{eqnarray}
Calculating the hyper-geometrical functions values in the right
part of Eq. \eqref{GrindEQ__39_} at the sound horizon we see these
functions have peculiarities at the sound horizon. Therefore the
auto-model solution satisfying the zero boundary conditions at the
sound horizon in the case under investigation is the trivial
solution  $C_{1} =C_{2} =0$ only. Summarizing this subsection let
us formulate the theorem

\noindent\textbf{Theorem.} \textit{There are no retarding
spherically-symmetric auto-model  solutions of the equation
\eqref{GrindEQ__1_} satisfying the zero boundary conditions
\eqref{GrindEQ__16_} at the sound horizon without the central
particle-like source ($\mu =0$).}

The proved theorem is analogues to the theorem about the
analytical solution of the Laplace equation in the spherical
symmetry case.

\section{Research of the auto-model solutions}
\subsection{Derivatives of the potential functions}

Let us come over to the analyses of the obtained auto-model
solutions in the particle-like source case. For that let us use
the expression \eqref{GrindEQ__8_} of the relative energy density
of perturbation and the expression \eqref{GrindEQ__9_} of the
radial medium velocity in the perturbation. At that we shall need
the expressions for the first and second derivatives of the
potential functions. Considering the definitions
\eqref{GrindEQ__11_} and \eqref{GrindEQ__12_} and the relations
\eqref{GrindEQ__20_} and \eqref{GrindEQ__27_} we shall get
expressions for the first and second radial derivatives of the
potential functions $\Psi (r,\eta )$

\begin{equation} \label{GrindEQ__40_}
\mathop{\Psi '}\nolimits_{r} (r,\eta )=\mu _{+} \eta ^{{\tfrac{1-3\kappa }{1+3\kappa }} }
\frac{2\Gamma \left({\tfrac{2}{1+3\kappa }} +{\tfrac{3}{2}} \right)}{\sqrt{\kappa \pi }
\Gamma \left({\tfrac{2}{1+3\kappa }} +1\right)} (1-z^{2} )^{{\tfrac{2}{1+3\kappa }} } ;
\end{equation}

\begin{equation} \label{GrindEQ__41_}
\quad \mathop{\Psi ''}\nolimits_{rr} (r,\eta )=-\frac{8\mu _{+}
\eta ^{-{\tfrac{6\kappa }{1+3\kappa }} } }{\sqrt{\pi } }
\frac{\Gamma \left({\tfrac{2}{1+3\kappa }} +{\tfrac{3}{2}} \right)}{\kappa (1+3\kappa )
\Gamma \left({\tfrac{2}{1+3\kappa }} +1\right)} (1-z^{2} )^{{\tfrac{1-3\kappa }{1+3\kappa }} } ,
\end{equation}
and for the temporal derivative of the function $\Phi (r,\eta )$ also

\begin{equation} \label{GrindEQ__42_}
\dot{\Phi }(r,\eta )=\eta ^{{\tfrac{1-3\kappa }{1+3\kappa }} } \left[2\frac{G(k,z)-\mu _{+} }{1+3\kappa } -\frac{2\mu _{+} \Gamma \left({\tfrac{2}{1+3\kappa }} +{\tfrac{3}{2}} \right)}{\sqrt{\pi } \Gamma \left({\tfrac{2}{1+3\kappa }} +1\right)} z(1-z^{2} )^{{\tfrac{2}{1+3\kappa }} } \right],
\end{equation}
here the function $G(\kappa ,z)$ is determined by the relation
\eqref{GrindEQ__24_} with the constant $C_{1} $ from
\eqref{GrindEQ__27_}.

From the given expressions it follows: by $\kappa >-1/3$ the first
radial and temporal derivatives of the potential functions $\Psi
(r,\eta )$ and $\Phi (r,\eta )$ turn into zero at the sound
horizon

\begin{equation} \label{GrindEQ__43_}
\mathop{\left. \frac{\partial }{\partial r} \Psi (r,\eta )\right|}\nolimits_{r=\sqrt{k} \eta } =\mathop{\left. \frac{\partial }{\partial r} \Phi (r,\eta )\right|}\nolimits_{r=\sqrt{k} \eta } =0;\quad (1+3\kappa \ge 0);
\end{equation}

\begin{equation} \label{GrindEQ__44_}
\mathop{\left. \frac{\partial }{\partial \eta } \Psi (r,\eta )\right|}\nolimits_{r=\sqrt{k} \eta } =\mathop{\left. \frac{\partial }{\partial \eta } \Phi (r,\eta )\right|}\nolimits_{r=\sqrt{k} \eta } =0;\quad (1+3\kappa \ge 0).
\end{equation}
At $\kappa <1/3$ the second radial derivatives of the potential functions turn into zero at the sound horizon

\begin{equation} \label{GrindEQ__45_}
\mathop{\left. \frac{\partial ^{2} }{\partial r^{2} } \Psi (r,\eta )\right|}\nolimits_{r=\sqrt{k} \eta } =\mathop{\left. \frac{\partial ^{2} }{\partial r^{2} } \Phi (r,\eta )\right|}\nolimits_{r=\sqrt{k} \eta } =0;\quad (1-3\kappa \ge 0).
\end{equation}
By $\kappa =1/3$ the second radial derivatives of the potential functions have a break of the first genus at the sound horizon. By  $\kappa >1/3$ the second radial derivatives of the potential functions  have a break of the second genus at the sound horizon.

\subsection{Evolution of the energy density distribution in
the spherical perturbation}
Calculating the relative density of the perturbation energy by the
formula \eqref{GrindEQ__8_} considering the relations
\eqref{GrindEQ__10_} and \eqref{GrindEQ__40_},\eqref{GrindEQ__41_}
we get finally
\[\frac{\delta \varepsilon }{\varepsilon _{0} } =-\frac{1}{z\eta \pi
\sqrt{\kappa } (1+3\kappa )}\times\]
\[ \left[3\frac{G(k,z)-\mu _{+}
}{1+3\kappa } -3\frac{\mu _{+} \Gamma \left({\tfrac{7+9\kappa
}{2(1+3\kappa )}} \right)}{\sqrt{\pi } \Gamma
\left({\tfrac{3(1+\kappa )}{1+3\kappa }} +1\right)} z(1-z^{2}
)^{{\tfrac{2}{1+3\kappa }} } \right. +\]
\begin{equation} \label{GrindEQ__46_}
+\left. \frac{2\mu _{+} \Gamma \left({\tfrac{7+9\kappa }{2(1+3\kappa )}} \right)}{\sqrt{\pi } \Gamma \left({\tfrac{3(1+\kappa )}{1+3\kappa }} +1\right)} (1-z^{2} )^{{\tfrac{1-3\kappa }{1+3\kappa }} } \right]\chi (1-z)=\frac{\mu _{+} }{\eta } \Delta (z)\chi (1-z),
\end{equation}
where the reduced relative density of the perturbation energy
$\Delta (z)$ is introduced. It can be strictly shown that $\Delta
(z)\ge 0.$

From this expression it is seen that by the time the profile form
of the perturbation energy density does not change relatively the
dimensionless radial variable $z=r/\sqrt{k} \eta $ and the
relative density of the perturbation energy decreases in inverse
proportion to the temporal variable $\eta $ (Fig.2).

At that in the terms of the common radial variable $r$ the energy
density perturbation profile is deformed. For example in Fig.3 the
evolution of the relative energy density perturbation profile at
the barotrope index $\kappa =1/6$ is shown. =3 Further on from the
formula \eqref{GrindEQ__46_} it is immediately seen that at
$\kappa <1/3$ the energy density perturbation at the sound horizon
vanishes, at $\kappa =1/3$ it has a final jump at the sound
horizon, and at $\kappa >1/3$ it has an infinite jump, it
absolutely corresponds to the behavior of the second radial
derivatives of the potential functions.

\begin{flushleft}
\begin{tabular}{ll}
\includegraphics[width=56mm, height=56mm]{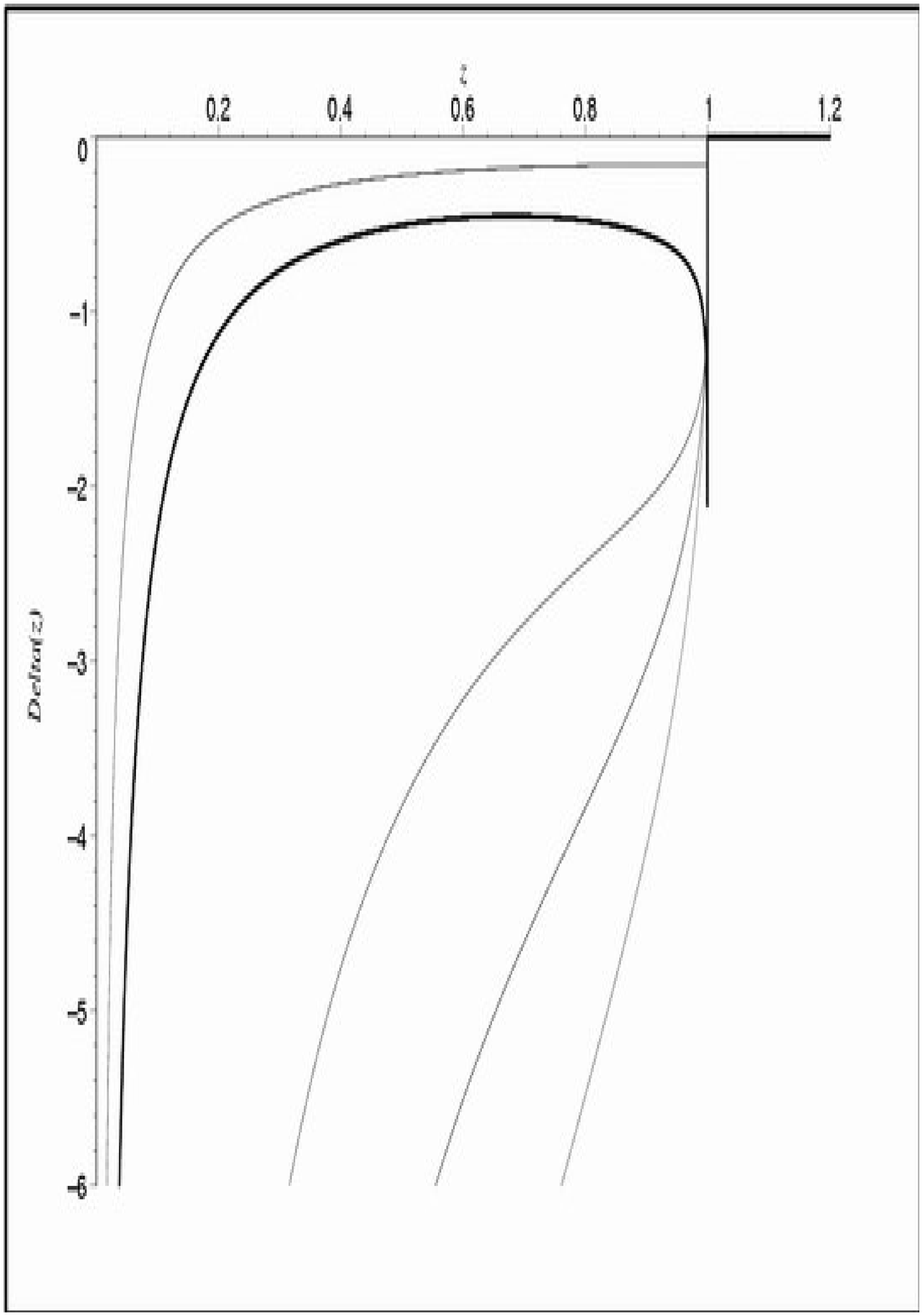}  &
\includegraphics[width=56mm, height=56mm]{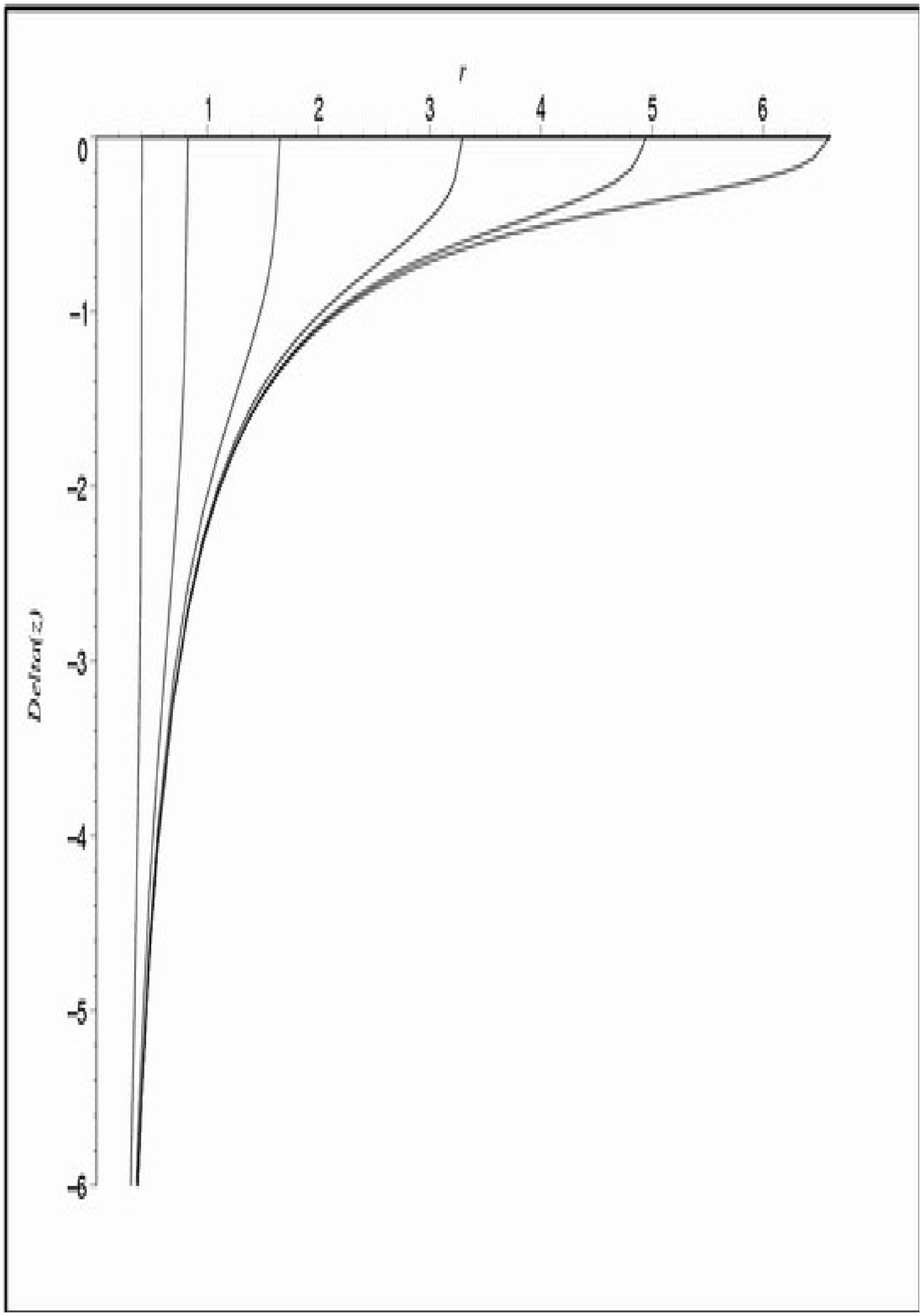} \\
\parbox{56mm}{\vskip 12pt\textbf{Fig. 2.} It is the evolution of the reduced relative
density of the perturbation energy  $\Delta $(z)  calculated by
the formula \eqref{GrindEQ__45_} as the function $z$. From bottom
up in fine line  $\kappa =1/6;1/5;1/4;1/3$; the heavy line
corresponds to $\kappa =1/2$.}   &
\parbox{56mm}{\vskip 2pt\textbf{Fig. 3.} It is the evolution of the
reduced relative density of the energy $\Delta $(z)  calculated by
the formula \eqref{GrindEQ__45_} as the function $r$ at $\kappa
=1/6$.  From the left to the right $\eta =1;2;4;8;12;16$.}
\end{tabular}
\end{flushleft}

\noindent

\noindent   Let us find out the  physical sense of the obtained solution. The energy perturbation corresponding to the nonsingular part of the potential function is described by the formula

\[\delta E=4\pi a^{3} \int _{0}^{\sqrt{\kappa } \eta } \, \delta \varepsilon r^{2} dr.\]
Hence considering \eqref{GrindEQ__10_} we get

\[\delta E=4\pi \eta ^{-{\tfrac{6\kappa }{1+3\kappa }} } \int _{0}^{\sqrt{\kappa } \eta } \, \frac{\delta \varepsilon }{\varepsilon _{0} } r^{2} dr.\]
Coming over to the dimensionless variable $z$ in the integral by the formula

\[r=\sqrt{\kappa } \eta z,\]
we bring it to the form

\begin{equation} \label{GrindEQ__47_}
\delta E=4\pi \mu _{+} \eta ^{{\tfrac{2}{1+3\kappa }} } \kappa ^{3/2} \int _{0}^{1} \, \Delta (z)dz\sim -m(\eta ),
\end{equation}
where (see (7)):

\begin{equation} \label{GrindEQ__48_}
m(\eta )=\mu _{+} \eta ^{{\tfrac{2}{1+3\kappa }} }
\end{equation}
is the mass of the central singular particle-like source. Thus, the full energy in the included in the nonsingular mode of the perturbation is negative and proportional to the mass of the particle-like source.

\subsection{Evolution of the fluid radial velocity in the
spherical perturbation}
Fulfilling the analogues calculations we
get  the expression for the radial velocity from
\eqref{GrindEQ__9_}

\[v=\frac{3}{8\pi \eta ^{2} \kappa ^{3/2} (1+3\kappa )} \left[\frac{G(k,z)-\mu _{+} }{z^{3} } \right. -
\frac{\mu _{+} \Gamma \left({\tfrac{7+9\kappa }{2(1+3\kappa )}} \right)}{\sqrt{\pi } \Gamma \left({\tfrac{3(1+\kappa )}{1+3\kappa }}
+1\right)} \frac{(1-z^{2} )^{{\tfrac{2}{1+3\kappa }} } }{z}\]
\begin{equation} \label{GrindEQ__49_}
\left.  -4\frac{\mu _{+} \Gamma \left({\tfrac{7+9\kappa
}{2(1+3\kappa )}} \right)}{\sqrt{\pi } \Gamma
\left({\tfrac{3(1+\kappa )}{1+3\kappa }} +1\right)} (1-z^{2}
)^{{\tfrac{1-3\kappa }{1+3\kappa }} } \right]=\frac{1}{\eta ^{2} }
\Upsilon (z).
\end{equation}
From this expression it is also seen that the radial velocity is
negative and its profile remains constant within the scale $z$,
and the absolute value of the velocity drops in inverse proportion
to $\eta ^{2} $. \vspace{12pt}

\begin{tabular}{l}
\includegraphics[width=100mm, height=80mm]{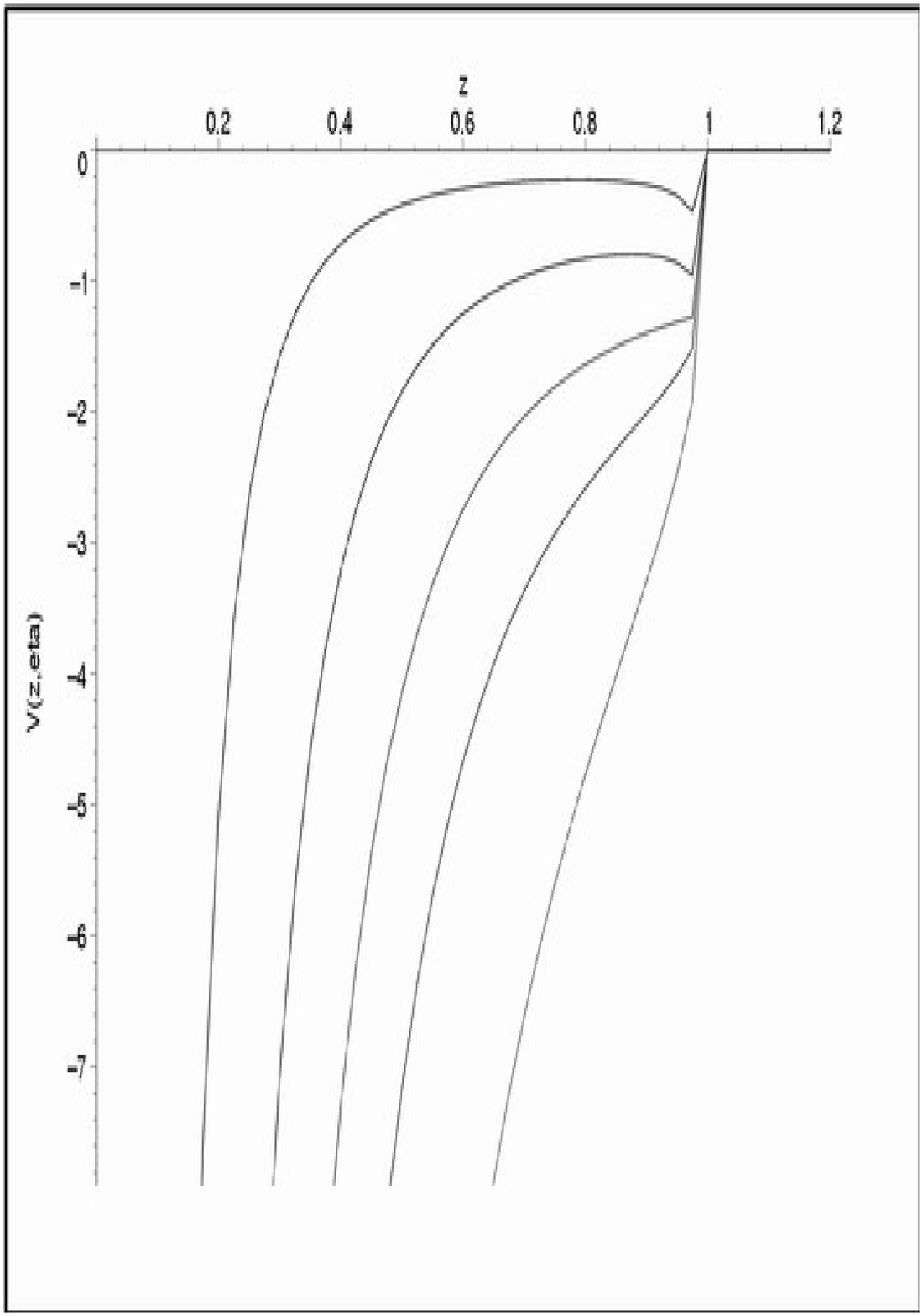}
\\[12pt]
\parbox{100mm}{\textbf{Fig. 4.} Dependence of the profile of the perturbation
radial velocity $\Upsilon (z)$ on the barotrop coefficient
at$\kappa =1/6$. From bottom up  $\kappa =1/6;1/4;1/3;1/2;1$ }\\
\end{tabular}


\begin{thebibliography}{99}
\bibitem{1}  Yu.G.Ignat'ev, N.Elmakhi. Izvestia Vuzov, Fizika, Vol. 41,
No 1, 2008, p. 66-76.
\bibitem{2}  Yu.G.Ignat'ev, N.Elmakhi. Izvestia Vuzov, Fizika,
Vol. 41, No 7,  2008, p. 69-76.
\bibitem{3}  Yu.G.Ignat'ev, .A.Popov.  ``Problems of the Theory of Gravitation,
Relativistic Kinetics and the Universe Evolution'', Kazan,
Izdatelstvo Kazan State Pedagogical University, 1988, p.5-16.
\bibitem{4}  Yu.G.Ignat'ev, A.A.Popov. Izvestia Vuzov, Fizika,
1989, No 5, p. 82-87
\bibitem{5}  Yu.G.Ignat'ev and A.A.Popov. Astrophysics and Space Science,1990, Vol 163, pp. 153-174.
\bibitem{6}  Yu.G.Ignat'ev, A.A.Popov. Physics Letters A, 1996,
Vol. 220, pp.22-29.
\bibitem{7}  N.N.Lebedev. Special Functions and their  applications, Moscow, GIPML, 1963.
\bibitem{8} A.P.Prudnikov, Yu.A.Brychkov, O.I.Marichev. Integrals and Series. Additional Chapters, Moscow, Nauka. 1986.
\end{thebibliography}
\end{document}